# Negative refraction and Negative refractive index in an optical uniaxial absorbent medium

Yi-Jun Jen,[1*] Ching-Wei Yu[1] and Chin-Te Lin[1]

[1]. Department of Electro-Optical Engineering, National Taipei University of Technology

## Abstract

This work demonstrates the existence of both negative refraction and a negative refractive index in an optical uniaxial absorbent medium that can be characterized by ordinary and extraordinary refractive indices. Negative refraction occurs in any absorbent uniaxial medium if the real part of the extraordinary index is less than its imaginary part. The refractive index is negative when the incident medium is sufficiently dense and the incident angle exceeds a critical angle that is defined here.

Negative refraction and a negative refractive index are extraordinary phenomena associated with the propagation of light in metamaterials (*1, 2*). An optical wave that is refracted by a metamaterial with a negative index of refraction would propagate in the direction of the negative refractive angle to satisfy Snell's law (*3*). Some materials with a positive real refractive index can also bend light toward the incident direction as a phenomenon of negative refraction (*4, 5*). This work demonstrates that both negative refraction and a negative refractive index can be found in an uniaxial absorbent medium that is found widely in nature.

The optical property of an uniaxial absorbent medium is characterized by its complex extraordinary principal index $\tilde{N}_e = n_e + i k_e$ and its ordinary principal index $\tilde{N}_o = n_o + i k_o$, which are defined according to an electric field that oscillates parallel and perpendicular to the optical axis, respectively. In this analysis, the reference coordinates (x, y, z) are coincident with the principal axes and the z-axis is the optical axis. The uniaxial medium is in the region $z > 0$. When a transverse magnetic (TM)-polarized wave is incident onto the medium, the penetrating wave becomes an extraordinary wave that experiences different principal indices in different directions of propagation. The direction of propagation of an electromagnetic wave is given by the Fresnel equation in terms of a complex unit wave vector (*6*) and principal refractive indexes

$$\frac{\tilde{s}_x^2}{\frac{1}{\tilde{N}_o^2} - \frac{1}{\tilde{N}^2}} + \frac{\tilde{s}_y^2}{\frac{1}{\tilde{N}_o^2} - \frac{1}{\tilde{N}^2}} + \frac{\tilde{s}_z^2}{\frac{1}{\tilde{N}_e^2} - \frac{1}{\tilde{N}^2}} = 0 \qquad (1)$$

where $\tilde{s}$ is the complex unitary wave vector and the complex refractive index $\tilde{N} = n + ik$ corresponds to the index experienced by the propagating wave. However, the actual wave vector that describes the propagation of the wave front in the anisotropic medium is the real part of $\tilde{N}\tilde{s}(\omega/c)$, where $\omega$ is the angular frequency and c is the velocity of light in free space. The equations of the boundary conditions for the electric field $\vec{E}$ and magnetic field $\vec{H}$ are solved to predict the energy flow, which is given the Poynting vector (ray vector) $\vec{P} = \vec{E} \times \vec{H}$. When the incident angle and wavelength of incident light are fixed, the time-varying ray vector varies in space and the tip of the time-varying vector follows an ellipse on the plane of incidence in the uniaxial medium (*7*). The angle of refraction of the time-averaged ray vector is given by a function of the incident angle and the refractive indices in the system. The calculation indicates that the refracted angle of the ray vector is negative provided the real part of the extraordinary index is less than its imaginary part. The condition for negative refraction is satisfied by various metal-dielectric composite materials. The proposed sample of silver nanowires embedded in an alumina matrix with complex ordinary index $\tilde{N}_o = 2.351 + i0.010$ and extraordinary index $\tilde{N}_e = 0.049 + i1.307$ at a wavelength of 633 nm is one such material (*5*). Let the incident medium be air. Figure 1A presents the real and imaginary parts of the refractive index $\tilde{N}$, as a function of the incident angle. The refracted angle of the ray vector varies from $0°$ to $-47.45°$ as the incident angle varies from $0°$ to $90°$, as presented in Fig. 1B.

Consider another condition, under which the incident medium is denser than the absorbent medium. When the refractive index of the incident medium exceeds an effective index of the medium $\tilde{N}$ that is a function of $\tilde{N}_o$ and $\tilde{N}_e$, the refracted angle of the wave vector exceeds the incident angle and a critical angle ($\theta_c = \sin^{-1}(\tilde{s}_x N_{eff} / N_0)$) can be defined as the refracted angle of the wave vector reaches $90°$. When the incident angle exceeds the critical angle, total reflection does not occur, but the index of refraction becomes negative. The real refraction index is negative and the component of the wave vector that is normal to the interface is inverted toward the incident medium. The wave vector and the ray vector point toward the incident medium and absorbent medium, respectively, which phenomenon is called a backward wave phenomenon. Effective medium theory *(8)* can be used to estimate the principal indices of the aforementioned sample at a wavelength of 365 nm: $\tilde{N}_o = 0.279 + i1.332$ and $\tilde{N}_e = 1.256 + i0.022$, yielding an effective index $N_{eff} = 1.144$. Let the incident medium $N_0$ be glass with refractive index 1.515: the critical angle is $52.91°$. The refraction index is negative when incident angles that exceed the critical angle, as presented in Fig. 1C. The negative refracted angle of wave vector varies from $-90°$ to $-59.63°$ as the incident angle varies from $52.91°$ to $90°$, as shown in Fig. 1D.

In conclusion, two conditions are given under which a uniaxial absorbent medium exhibits negative refraction and has a negative refractive index, respectively. People can easily observe the two extraordinary phenomena by manufacturing or searching such an anisotropic medium in nature. Additionally, the energy flow and propagation of wave can be separately manipulated using the anisotropic absorbent property instead of complex artificial metamaterial.

**References and Notes**

1. D. R. Smith, J. B. Pendry, M. C. K. Wiltshire, *Science* **305**, 788 (2004).

2. J. B. Pendry, *Phys. Rev. Lett.* **85**, 3966 (2000).

3. M. Born, E. Wolf, *Principles of Optics* (Cambridge Univ. Press, New York, ed. 7, 1999), pp. 38-40.

4. J. Yao *et al.*, *Science* **321**, 930 (2008).

5. Y. Liu, G. Bartal, X. Zhang, *Opt. Express* **16**, 15439 (2008).

6. J. M. Diñeiro *et al.*, *J. Opt. Soc. Am. A* **24**, 1767 (2007).

7. More detals are available as supporting material on Science Online.

8. G. W. Milton, *J. Appl. Phys.* **52**, 5286 (1981).

9. The authors would like to thank the National Science Council of the Republic of China, Taiwan, for financially supporting this research under Contract No. NSC 96-2221-E-027-051-MY3.

**Figure caption**

Fig. 1 (A) Real part and imaginary part of refractive index $\tilde{N}$. (B) Refracted angles of wave vector and ray vector as functions of incident angle for uniaxial medium with complex ordinary index $\tilde{N}_o = 2.351 + i0.010$ and extraordinary index $\tilde{N}_e = 0.049 + i1.307$ at wavelength of 633 nm.

(C) Real part and imaginary part of refractive index $\tilde{N}$. (D) Refracted angles of wave vector and ray vector as functions of incident angle for an uniaxial medium with complex ordinary index $\tilde{N}_o = 0.279 + i1.332$ and extraordinary index $\tilde{N}_e = 1.256 + i0.022$ at wavelength of 365 nm.

Fig. 1

Supporting Online Material for

# Negative refraction and Negative refractive index in an optical uniaxial absorbent medium

Yi-Jun Jen, [1*] Ching-Wei Yu[1] and Chin-Te Lin[1]

[1]. Department of Electro-Optical Engineering, National Taipei University of Technology

[*] To whom correspondence should be addressed. E-mail: jyjun@ntut.edu.tw

## Supporting Online Material

As shown in Fig. 1, the reference coordinates are (x, y, z) with the z axis parallel to the optical axis of the uniaxial medium. The principal refractive indices are $\tilde{N}_o = \tilde{N}_x = \tilde{N}_y$ and $\tilde{N}_e = \tilde{N}_z$. The oscillating electric and magnetic fields are given by Eq. (*S1*)

$$\tilde{F} = F\tilde{u}_F \exp\left[-i\omega\left(t - \frac{\tilde{N}}{c}\vec{r}\cdot\tilde{s}\right)\right] \qquad (S1)$$

where $\omega$ is the angular frequency; c is the velocity of light in free space; F is the field amplitude, and $\vec{r}$ is the position vector $x\hat{e}_x + z\hat{e}_z$. The complex unitary filed vector $\tilde{u}_F$, the refractive index $\tilde{N}$ and the complex unitary wave vector $\tilde{s}$ are complex. The complex refractive index $\tilde{N}$ that is experienced by the propagating wave experiences satisfies the Fresnel equation,

$$\frac{1}{\tilde{N}^2} = \frac{\tilde{s}_x^2}{\tilde{N}_e^2} + \frac{\tilde{s}_z^2}{\tilde{N}_o^2} \qquad (S2)$$

From Eq. (S1), the actual wave vector $\vec{k}$, which describes the propagation of the wave front is

$$\vec{k} = \text{Re}\left\{\frac{\omega}{c}\tilde{N}\tilde{s}\right\} \qquad (S3)$$

Figure 1A plots the the real and imaginary parts of the extraordinary refractive index $\tilde{N}$ for the of the air/anisotropic medium system with anisotropic optical constants $n_o + ik_o$ and $n_e + ik_e$ )

A transverse magnetic (TM)-polarized wave that propagates in the uniaxial medium is an extraordinary wave. Diñeiro *et al.* described the associated extraordinary electric field, the electric displacement and the magnetic field (*S2*).

$$\tilde{E} = E\left(\frac{\tilde{s}_z}{\tilde{N}_o^2}\hat{e}_x - \frac{\tilde{s}_x}{\tilde{N}_e^2}\hat{e}_z\right)\exp\left[-i\omega\left(t - \frac{\tilde{N}}{c}\vec{r}\cdot\tilde{s}\right)\right] \qquad (S4)$$

$$\tilde{D} = \varepsilon_0 E(\tilde{s}_z\hat{e}_x - \tilde{s}_x\hat{e}_z)\exp\left[-i\omega\left(t - \frac{\tilde{N}}{c}\vec{r}\cdot\tilde{s}\right)\right] \qquad (S5)$$

$$\tilde{H} = \frac{\varepsilon_0 cE}{\tilde{N}}\hat{e}_y\exp\left[-i\omega\left(t - \frac{\tilde{N}}{c}\vec{r}\cdot\tilde{s}\right)\right] \qquad (S6)$$

where $\varepsilon_0$ is the permittivity of free space, and E is the amplitude of the electric field. The time-varying extraordinary Poynting vector is calculated as,

$$\frac{1}{2}\text{Re}\{\tilde{E}\times\tilde{H}\} = \frac{\varepsilon_0 cE^2}{2}\left[\left|\frac{\tilde{s}_x}{\tilde{N}_e^2\tilde{N}}\right|\cos(2\theta_p + \theta_1)\hat{e}_x + \left|\frac{\tilde{s}_z}{\tilde{N}_o^2\tilde{N}}\right|\cos(2\theta_p + \theta_2)\hat{e}_z\right] \qquad (S7)$$

where

$$\theta_p = \text{Re}\left\{-\omega\left(t - \frac{\tilde{N}}{c}\vec{r}\cdot\tilde{s}\right)\right\} \qquad (S8)$$

$$\theta_1 = \arg\left(\frac{\tilde{s}_x}{\tilde{N}_e^2\tilde{N}}\right) \qquad (S9)$$

$$\theta_2 = \arg\left(\frac{\tilde{s}_z}{\tilde{N}_o^2\tilde{N}}\right) \qquad (S10)$$

locus of the tip of the time-varying Poynting vector is an ellipse, called the Poynting ellipse (*S3*). The time-averaged Poynting vector is given by,

$$\vec{P}_{av} = \frac{1}{2} Re\left\{\tilde{E}^* \times \tilde{H}\right\} = \frac{\varepsilon_0 cE^2}{2} Re\left\{\left(\frac{\tilde{s}_x}{\tilde{N}_e^2}\right)^*\left(\frac{1}{\tilde{N}}\right)\hat{e}_x + \left(\frac{\tilde{s}_z}{\tilde{N}_o^2}\right)^*\left(\frac{1}{\tilde{N}}\right)\hat{e}_z\right\} \qquad (S11)$$

The parallel-to-interface component of the vector $P_x$ is further expanded by substituting the unitary wave vector $\tilde{s}_x$ for refractive index of the incident medium $N_0$ and incident angle $\theta_0$.

$$\vec{P}_{av} \cdot \hat{e}_x = Re\left\{\frac{\varepsilon_0 cE^2}{2}\left(\frac{\tilde{s}_x}{\tilde{N}_e^2}\right)^*\left(\frac{1}{\tilde{N}}\right)\right\} = \frac{\varepsilon_0 cE^2}{2} \frac{N_0 \sin\theta_0}{\left|\tilde{N}\right|^2} \frac{Re[\tilde{N}_e]^2 - Im[\tilde{N}_e]^2}{(Re[\tilde{N}_e]^2 + Im[\tilde{N}_e]^2)^2} \quad (S12)$$

From Eq. (S12), $P_x$ is negative if the magnitude of the real extraordinary index is less than the imaginary extraordinary index.

The critical angle of the wave vector is defined by comparing the wave vector in the incident medium with that in the uniaxial medium. The critical angle $\theta_c$ satisfies the equality between the parallel-to-interface components of the wave vector in the uniaxial absorbent and the incident medium.

$$\frac{\omega}{c} N_{eff} \tilde{s} \cdot \hat{e}_x = \frac{\omega}{c} N_0 \sin\theta_c \qquad (S13)$$

When the incident angle exceeds the critical angle, the real part of the effective index, given by the Fresnel equation, is negative. The actual wave vector in the crystal comprises a negative z component and a positive x component. The critical angle is presented as a function of the optical constants.

$$\theta_c = \sin^{-1}\left(\sqrt{\frac{1}{(\alpha\beta - \gamma)N_0^2}}\right) \qquad (S14)$$

where $\alpha$, $\beta$ and $\gamma$ are functions of the principal refractive indices.

$$\alpha = \frac{\text{Re}[\tilde{N}_o]^2 - \text{Im}[\tilde{N}_o]^2}{2\,\text{Re}[\tilde{N}_o]\,\text{Im}[\tilde{N}_o]} \tag{S15}$$

$$\beta = 2\left(\frac{\text{Re}[\tilde{N}_o]\,\text{Im}[\tilde{N}_o]}{(\tilde{N}_o \cdot \tilde{N}_o^*)^2} - \frac{\text{Re}[\tilde{N}_e]\,\text{Im}[\tilde{N}_e]}{(\tilde{N}_e \cdot \tilde{N}_e^*)^2}\right) \tag{S16}$$

$$\gamma = \frac{\text{Im}[\tilde{N}_e]^2 - \text{Re}[\tilde{N}_e]^2}{(\tilde{N}_e \cdot \tilde{N}_e^*)^2} + \frac{\text{Re}[\tilde{N}_o]^2 - \text{Im}[\tilde{N}_o]^2}{(\tilde{N}_o \cdot \tilde{N}_o^*)^2} \tag{S17}$$

The condition for the negative refractive index is that the incident medium, with index $N_0$, has to be sufficiently dense to satisfy the inequality,

$$N_0 > \sqrt{\frac{1}{\alpha\beta - \gamma}} = N_{\text{eff}} \tag{S18}$$

The effective index $N_{\text{eff}}$ is given by functions $\alpha$, $\beta$ and $\gamma$, which must satisfy the inequality.

$$(\alpha\beta - \gamma) > 0 \tag{S19}$$

Under these conditions, the negative refractive index is observed when the incident angle exceeds the critical angle.

## References

S1. S. Alfonso *et al.*, *J. Opt. Soc. Am. A* **21**, 1776 (2004).

S2. J. M. Diñeiro *et al.*, *J. Opt. Soc. Am. A* **24**, 1767 (2007).

S3. P. Halevi, A. Mendoza-Hernández, *J. Opt. Soc. Am.* **71**, 1238 (1981).